\documentclass[aps,pra,twocolumn,floatfix,showpacs]{revtex4-1}
\usepackage[utf8]{inputenc}
\usepackage{graphicx}
\usepackage{times}
\usepackage{amsmath}
\usepackage{amsfonts}
\usepackage{braket}
\usepackage{subfigure}
\usepackage{hyperref}
\usepackage{array}
\newcommand{\expected}[1]{\left\langle #1\right\rangle}
\newcommand{\id}{\mathbb{I}}
\DeclareMathOperator{\Tr}{Tr}
\newcommand{\hamiltonian}{\mathcal{H}}
\newcommand{\rtn}{\text{\tiny RTN}}
\newcommand{\ou}{\text{\tiny OU}}
\newcommand{\negativity}{\mathcal{E}}

\newcommand{\ie}{\text{\tiny{IE}}}
\newcommand{\ce}{\text{\tiny{CE}}}
\newcommand{\rhp}{\mathcal{N}_\text{\tiny RHP}}
\newcommand{\blp}{\mathcal{N}_\text{\tiny BLP}}
\begin{document}
\title{Non-Markovian dynamics of single- and two-qubit systems 
interacting with Gaussian and non-Gaussian fluctuating transverse 
environments}
\author{Matteo A. C. Rossi}
\email{matteo.rossi@unimi.it}
\homepage{http://users.unimi.it/aqm}
\affiliation{Quantum Technology Lab, 
Dipartimento di Fisica, Universit\`a degli Studi di Milano, 
20133 Milano, Italy.}
\author{Matteo G. A. Paris}
\email{matteo.paris@fisica.unimi.it}
\homepage{http://users.unimi.it/aqm}
\affiliation{Quantum Technology Lab, Dipartimento di Fisica, 
Universit\`a degli Studi di Milano, 20133 Milano, Italy.}
\affiliation{CNISM, Unit\`a Milano Statale, I-20133 Milano, Italy.}
\date{\today}
\begin{abstract}
We address the interaction of single- and two-qubit systems with
an external {\em transverse} fluctuating field and analyze in details the
dynamical decoherence induced by Gaussian and non-Gaussian noise, e.g.
random telegraph noise (RTN). Upon exploiting the exact RTN solution of
the time-dependent Von Neumann equation, we analyze in details the
behavior of quantum correlations and prove the non-Markovianity of the
dynamical map in the full parameter range, i.e. for either fast or slow
noise. The dynamics induced by Gaussian noise is studied numerically and
compared to the RTN solution, showing the existence of (state dependent) 
regions of the parameter space where the two noises lead to very similar
dynamics. Our results shows that while the effects of non-Gaussian noise
cannot be trivially mapped to that of Gaussian noise and viceversa, i.e.
the spectrum alone is not enough to summarize the noise effects, the 
dynamics under the effect of one kind of noise may be {\em
simulated} with high fidelity by the other one.
\end{abstract}
\maketitle
%%%%%	
\section{Introduction}
\label{sec:introduction}
The unavoidable interaction of a quantum system with its environment
generally causes decoherence and a loss of quantumness, thus posing a
threat to quantum information processing. A deep understanding of the
decoherence mechanisms in quantum systems, together with the capability
to engineer the environment, are thus very important steps toward the
development of quantum technologies.
\par
In general, a quantum system interacts with a complex environment that
should be described quantum-mechanically. This is often challenging or
even unfeasible in practice, unless one recurs to perturbative
approximations \cite{yan05,priv03} or to approximations that reduce the description
of the environment to a few degrees of freedom \cite{
Leg87,laikh95,ning13,dvira14}. In many
situations, the environment may be conveniently represented as a
collection of fluctuators, such that it can be described as a classical
stochastic field such as, for instance, a Gaussian process or random
telegraph noise (RTN) \cite{xu03,Paladino2014}. A relevant example
is that of charge noise in superconducting qubits or quantum dots, 
which may be conveniently modeled by a classical field as far as 
the charge fluctuators couple more strongly to their own
environment than to the qubit \cite{Paladino2002,Galperin2006,ramon15}. 
In other regimes, solid state 
or superconducting devices can be conveniently described 
by models in which noise is due to a collection of bistable fluctuators, 
resulting in a $1/f$ spectrum, $f$ being the frequency. 
\par
It is subject of current research \cite{lacroix08,Helm2009,Helm2011,Crow2014,witzel14} whether
the interaction with quantum environments may be effectively described
by a classical stochastic field. So far, full equivalence has been shown
in \cite{Crow2014} for single-qubit dephasing, with an explicit
construction of the probability distribution required for the classical
stochastic process to describe the quantum environment. General 
arguments valid for RTN noise have been also discussed \cite{saira07}.
A stochastic process approach may be also used to decouple the dynamics of
the system from that of its environment, with the two separated systems 
evolving in common classical random fields \cite{shao04}.
\par
Among the different classes of open quantum systems, a large attention
has been put to qubit systems subject to environmental noise inducing a
dephasing dynamics \cite{Paladino2014,Rossi2014b}, i.e. noise with
typical frequencies that are smaller than the characteristic frequencies
of the quantum system. In these situations, the energy of the system is
not altered by the interaction and only the coherences are affected. For
the dephasing model, analytic solutions have been found for Gaussian
noise \cite{Yu2006} and RTN \cite{Bergli2009}, and numerically for colored noise
\cite{Benedetti2013a}. A number of interesting features have been
discovered and studied, such as entanglement sudden death (ESD)
\cite{Yu2006} and quantum discord freezing \cite{Mazzola2011}. Moreover
the non-Markovianity of the dynamics has been addressed
\cite{Benedetti2014c}, and the use of qubits as probes for the
spectral properties of the environment has been proposed
\cite{Benedetti2014a,Benedetti2014b,Alvarez2011,Zwick2015}. Recently, the role of
entanglement in improving the estimation of dephasing environments 
has been recognized \cite{Rossi2015}, thus making of interest the study
of decoherence in more general environments \cite{Szank15,Szank15a}.
%%%
\par
The dephasing Hamiltonian for a single qubit under the effect of an
external field is
$\hamiltonian(t) = \omega \sigma_z + \lambda\, B(t)\,\sigma_z$, 
where $\omega$ is the energy between the energy levels, $\lambda$ is a
coupling constant, and $B(t)$ is a stochastic process that models the
external noise (we set $\hbar=1$). This is often referred to as
{\em longitudinal noise}, the direction of the external driving 
being parallel to the qubit axis in the spin space. In turn, in 
a dephasing model, the populations of the system are constant.
\par
If the typical frequencies of the environment are close to the
characteristic frequency of the qubit, the interaction induces
transitions between the energy levels and the pure dephasing model is
inadequate to describe the dynamics. The Hamiltonian must include a
transverse interaction \cite{Vestgarden2008,Lombardo2014,Crescimanno1993} and, in general, may be written as
\begin{equation}\label{eq:1_qubit_hamiltonian_general}
	\hamiltonian(t) = \omega \sigma_z +  B(t)\, 
	\boldsymbol{\lambda} \cdot \boldsymbol \sigma
\end{equation}
\par
In this paper, we analyze in details the case in which the interaction
is purely transverse, i.e. when $\lambda_z = 0$. We address the dynamics
of single- and two-qubit systems under the effect of RTN, a non-Gaussian
kind of noise, and provide an exact analytic description of the
resulting decoherence process. We also address numerically the dynamics
induced by Gaussian noise.  We analyze the evolution of quantum
correlations, evaluate the non-Markovianity of the dynamical map, and
compare the effects of the two kinds of noise, looking  for features
that depends on the sole spectrum of the noise rather than its
statistics. Our results shows that the effects of non-Gaussian noise
cannot be trivially mapped to that of Gaussian noise and viceversa, i.e.
the spectrum alone is not enough to summarize the noise effects. 
On the other hand, the dynamics under the effect of one kind of noise 
may be effectively simulated, i.e. with high fidelity, by the other one
with a suitable choice of the noise parameters.
\par
Besides, we have identified, for both kind of noise, two different 
working regimes. In the first one, when the spectral width of the noise 
$\gamma$ is small, quantum correlations oscillate
heavily and there are sudden deaths and rebirths of entanglement. 
The frequency of oscillations depends on $\omega$
and is doubled if the two qubits are affected by a common environment.
In the second regime, the correlations decay to zero, with sudden death
of entanglement and with oscillations. The time constant of the decay is
roughly inversely proportional to $\gamma$, i.e. the decay is slower for
very fast noise. The different features of the dynamics, however, cannot be
linked to a transition in the structure of the dynamical map, which is
non-Markovian in the full parameter range, i.e. for either fast 
or slow noise. 
\par
The structure of the paper is as follows: in Section~\ref{sec:model} we
present the model and introduce the measures of quantum correlations and
non-Markovianity. In Section~\ref{sec:dynamics} we present the solution
of the dynamics of the system, whereas in Section~\ref{sec:comparison} 
we study the evolution of quantum correlations and compare the 
dynamics induced by the two kinds of noise. In Section~\ref{sec:comparison_of_non_markovianity_measures} we discuss the
non-Markovianity of the dynamical maps, whereas Section
\ref{sec:conclusions}
closes the paper with some concluding remarks.
%%%
\section{The model} 
\label{sec:model}
We consider a qubit characterized by the energy splitting $\omega$, and
affected by a transverse noise. The Hamiltonian is
\begin{equation}\label{eq:1_qubit_hamiltonian}
	\hamiltonian(t) = \omega \sigma_z + \lambda B(t)\sigma_x,
\end{equation}
where we assume without loss of generality that the noise acts in the
$x$ direction. The evolution operator for the Hamiltonian in Eq.
\eqref{eq:1_qubit_hamiltonian}, for a given realization of the
stochastic process $B(t)$, is
\begin{equation}\label{eq:evolution_operator}
	U(t) = \mathcal T \exp \left(-i \int_0^t \hamiltonian(t')dt' \right),
\end{equation}
where $\mathcal T$ is the time-ordering operator, which is required
because the Hamiltonian doesn't commute with itself at different times.
If the qubit is initially prepared in the state described by the density
matrix $\rho_0$, the density matrix at the time $t$ is 
\begin{equation}
\label{eq:evolved_state}
	\rho(t) = \expected{U(t)\rho_0 U(t)^\dagger},
\end{equation} 
where $\expected{\cdot}$ denotes the average over all
possible realizations of the stochastic process $B(t)$. 
Equation \eqref{eq:evolved_state} describes a convex combination of
unitary operators, which itself provides the Kraus decomposition of the
corresponding CPT map.
\par
We are also going to consider a system of two identical, non-interacting
qubits each interacting with a noisy environment, in order to study the
evolution of quantum correlations between the qubits. The two-qubit
Hamiltonian reads 
\begin{equation}
\hamiltonian(t) = 
\hamiltonian_1(t) \otimes \id_2 + \id_1 \otimes \hamiltonian_2(t),
\end{equation}
where $\hamiltonian_i(t)$ have the form of Eq.
\eqref{eq:1_qubit_hamiltonian} and the $B_i(t)$ may be correlated (if
the two qubits interact with a common environment) or completely
uncorrelated (in the case in which the two qubits are affected by
independent environments, IE). For simplicity, we'll consider $B_1(t) =
B_2(t)$ in the common environment (CE) case.
\par
A Gaussian process is fully characterized by its second
order statistics, i.e. by its mean $\mu$ and its autocorrelation
function $K$, in formula \begin{align}	
\mu(t) &= \expected{B(t)} \\
K(t,t') &= \expected{B(t)B(t')}.
\end{align}
In this work, we employ the Ornstein-Uhlenbeck (OU) process
\cite{DeChiara2003,Fiasconaro2009,Benedetti2014d} as a paradigmatic
stationary stochastic process with finite-time correlations. We set
$\mu(t)\equiv 0$ and assume the following autocorrelation function:
\begin{equation}
K_\ou(t-t') = e^{-2 \gamma |t-t'|},
\end{equation}
which corresponds to a Lorentzian spectrum 
\begin{equation}\label{eq:spectrum}
S(\omega) = \frac{4 \gamma}{4 \gamma ^2+\omega ^2}.
\end{equation}
with spectral width $2\gamma$. For $\gamma \rightarrow
\infty$, $K(t-t')\sim \delta(t-t')$, i.e. the OU process reduces to
white noise.
\par
RTN noise is produced by bistable fluctuators, i.e. systems where 
a quantity flips between two values with a certain switching rate, 
such as a resistance switching between two discrete values, charges 
jumping between two different locations, or electrons that flip their 
spin. In order to describe classical environment inducing RTN, 
the quantity $B(t)$ in Eq. \eqref{eq:1_qubit_hamiltonian} should flip 
randomly between the values $\pm 1$ with a given switching rate 
$\gamma$. This kind of noise is also characterized by an 
exponentially decaying autocorrelation function 
\begin{equation}
K_{\rtn}(t-t') = e^{-2\gamma|t-t'|}
\end{equation} 
and by a Lorentzian spectrum $S(\omega)$, Eq. \eqref{eq:spectrum}, i.e.
the  OU and RTN process have exactly the same autocorrelation function. 
The latter, however, being a non-Gaussian process, cannot be fully described
by means of its first and second moments.
\par
For either kind of noise, the model exhibits a natural scaling 
property in terms of the coupling, which may be exploited in order 
to work with dimensionless quantities. Indeed, we rescale all the
quantities in terms of the coupling $\lambda$ by performing the
following substitutions 
$$t \rightarrow \lambda t, \quad \gamma \rightarrow
\gamma / \lambda, \quad \omega \rightarrow \omega/\lambda\,.$$ 
The
Hamiltonian, Eq. \eqref{eq:1_qubit_hamiltonian}, now reads
\begin{equation}
	\hamiltonian(t) = \omega \sigma_z + B(t)\sigma_x.
\end{equation}
%%%%
\subsection{Quantum correlations} % (fold)
\label{sub:quantum_correlations}
In the following we will study the dynamics of quantum correlations by 
evaluating negativity \cite{Vidal2002} as a measure of entanglement 
and using entropy \cite{Ollivier2001} to define quantum discord. 
Negativity is defined as
\begin{equation}\label{eq:negativity}
	\negativity = 2 \left|\sum_i \lambda_i^-\right|,
\end{equation}
where $\lambda_i^-$ are the negative eigenvalues of the partial
transpose of the density matrix with respect to either of the qubits. We
remark that the negativity of the partial transpose is necessary and
sufficient for two-qubit systems to be entangled.
\par
Quantum discord is defined as the difference between the total
correlations and the classical correlations between the two subsystems:
\begin{equation}\label{eq:discord}
	\mathcal{D} = \mathcal{I} - \mathcal{C}.
\end{equation}
Total correlations are given by the quantum mutual information
$\mathcal{I} = S(\rho_A) + S(\rho_B) - S(\rho)$, where $S$ is the Von
Neumann entropy, and $\rho_A$ and $\rho_B$ are the reduced density
matrices of the two subsystems. Classical correlations, induced by a
measurement on one of the two subsystems, are given by $\mathcal{C} =
\max_{\{B_k\}} [S(\rho_A) - S(\rho|\{B_k\})]$, where $S(\rho|\{B_k\})$ is
the conditional entropy of the state of the two-qubit system with
respect to the outcome of the measurement $\{B_k\}$ on system $B$, and
the maximization is carried over all possible projective measurements.
\par
The evaluation of quantum discord is in general a difficult task, as it
involves an optimization procedure. For two-qubit systems, an analytic
result was found by Luo \cite{Luo2008} for a subset of the state space,
i.e. for those states that have maximally mixed marginals. As we are
going to show below, if the initial state of the system belongs to this
subset, the dynamics induced by transverse noise is limited to this subset,
so we will employ Luo's formula in the following.
\subsection{Non-Markovianity measures} % (fold)
\label{sub:non_markovianity_measures}
The concept of non-Markovianity for quantum dynamical maps is related to
the concept of divisibility, i.e. if $\mathcal{E}(t_2,t_0)$ is the
operator describing the quantum map from time $t_0$ to $t_2$, the map is
divisible if it is completely positive (CP) and 
\begin{equation}
\mathcal{E}(t_2,t_0) = \mathcal{E}(t_2,t_1)\mathcal{E}(t_1,t_0)
\end{equation}  
for every intermediate time $t_0 < t_1 < t_2$.
We characterize the non-Markovianity of the quantum map by considering
two measures: the entanglement-based RHP measure \cite{Rivas2010} and
the BLP measure \cite{Breuer2009}, based on the time evolution of the
trace distance. These two measures define sufficient conditions for the
dynamical map to be non-Markovian. Here we briefly review the two 
measures.
\paragraph{RHP measure} % (fold)
\label{par:rhp_measure}
We consider the quantum system of interest to be in the maximally
entangled state 
\begin{equation} \ket\psi = \frac{1}{\sqrt{N}}
\sum_{n=1}^N\ket{n}_S\ket{n}_A,
\end{equation}
where $\ket{n}$ are the vectors of a basis of the Hilbert space of the
system. We now let the system $S$ interact with the environment and
evaluate the entanglement of the state $\ket{\psi(t)}$. Since any
entanglement measure is a monotone under local CP maps, any increase of
an entanglement measure with time denotes that the dynamical map fails
to be divisible, i.e. that it is non-Markovian. The RHP is defined
quantitatively as \begin{equation}
\rhp = \int_{t_0}^{t_f} \left| \frac{dE(t)}{dt} \right|,
\end{equation}
where $E(t)$ is any entanglement measure (in our case, the negativity).
In fact, Ref. \cite{Rivas2010} introduces another measure that is a
necessary and sufficient condition for the non-Markovianity of the
quantum map, based on the Choi-Jamiolkowski isomorphism. However, to
compute this measure one needs to know the structural form of the
dynamical map between any two time instants, which is not the case for
our processes.
\paragraph{BLP measure} % (fold)
\label{par:blp_measure} The BLP measure is based on the fact that the
trace distance, $D(\rho_1,\rho_2) = \frac 12
\Tr\left[\sqrt{(\rho_1-\rho_2)^2}\right]$, is contractive for CP maps,
so if the quantum map is divisible, then for any pair of initial states
of the system the trace distance between the evolved states is a
monotonically decreasing function of time. If in a certain time interval
the trace distance increases, we can say that the CP map under
investigation is non-Markovian, because the map fails to be divisible in
that interval of time.The BLP measure is computed by integrating with
respect to time the positive part of the time derivative of the trace
distance and then optimizing the result over all possible pairs of
states: 
\begin{equation}\label{eq:blp}
\blp = \max_{(\rho_1,\rho_2)} \int_{t_0}^{t_f}  
\left[\frac{d}{dt}D(t,\rho_1,\rho_2)\right]_+dt.
\end{equation}
Calculating the BLP measure may be challenging in general, as the optimization
over all possible pairs of states is required. For qubits however, the
optimization can be restricted to the surface of the Bloch sphere
\cite{Wissmann2012}, leaving only the polar and azimuth 
angles as parameters to optimize over.
%%%
\section{\label{sec:dynamics}Solution of the dynamics} % (fold)
In order to obtain a solution for $\rho(t)$ in Eq. \eqref{eq:evolved_state} 
one should at first find an explicit expression for the evolution 
operator $U(t)$ in Eq. \eqref{eq:evolution_operator} and then 
calculate the expected value over all possible realizations of the 
stochastic process. 
%%%
\subsection{Gaussian noise: numerical simulation} % (fold)
\label{sub:numerical_simulation}
For Gaussian noise an explicit expression for $U(t)$ is only possible by
means of approximations such as the Dyson series or the Magnus expansion
\cite{Magnus1954}, which are valid in a neighborhood of the initial
time. A cumulant expansion has been also introduced and discussed in the 
single-qubit case \cite{skinner87,aihara90,der91,andreozzi92}. 
The lack of an analytic solution 
is due to the fact that the time-dependent Hamiltonian does
not commute with itself at different times and we cannot find an explicit
expression for the time-ordered exponential. An analytic result can be
obtained in the approximation of a quasi-static external field, i.e.
when the stochastic process is weakly dependent on time and the two-time
commutator for the Hamiltonian is negligible \cite{Benedetti2014d}.
\par
The dynamics of the system may be studied numerically using different 
approaches \cite{aihara90,risken90}. 
We proceed in a straightforward way by numerical evaluation
of the unitary propagator upon 
discretizing the time interval $[0,t]$ in $n$ steps of length 
$\Delta t$. $\Delta t$ should be
small enough for $H(t)$ to be approximately constant in the time
interval. The evolution operator from $t_i$ to $t_{i+1}$ for a specific
realization of the process $B(t)$ reads $U_{t_i,t_{i+1}} \simeq \exp[-i
H(t_i)\Delta t]$. The density operator of the qubit is then given by
\begin{equation} 
\rho(t) \simeq \expected{ U_{t_{n-1},t_n}\cdots
U_{t_1,t_2} \rho_0 U^\dagger_{t_1,t_2} \cdots U^\dagger_{t_{n-1},t_n}}.
\end{equation}
The expected value is obtained from a sufficiently large number $N$ of
randomly generated realizations of the noise. This method converges as
$N$ increases. We have checked that the standard deviation decreases
as $1/\sqrt{N}$. Typical values for $N$ are of the order of $10^{5}$ to
$10^{6}$, with the maximum relative error on an element of the density
matrix of the order of $10^{-3} \div 10^{-4}$ after $100$ evolution
steps.
%%%
\subsection{Analytic solution for the RTN} % (fold)
\label{sub:analytical_solution_for_the_rtn} 
Analytic solutions for a qubit interacting with RTN with an arbitrary
direction are known \cite{Cheng2008,Goychuk1995,Goychuk2006}. By following \cite{Cheng2008}, we consider the
time evolution of the Bloch vector $\mathbf{n}(t)$, which can be written by means of a transfer matrix $T$ applied to the initial Bloch vector $\mathbf{n}(0)$ as 
\begin{equation}
\mathbf{n}(t) = T \mathbf{n}(0) = \expected{T_{s_n}
\cdots T_{s_1} }\mathbf{n}(0),
\end{equation}
where $T_{s_i}$ is the  $3\times 3$ transfer matrix from the time
instant $t_i$ to time $t_{i+1}$, when the fluctuator is in the state
$s_i = \pm 1$. $T_{s_i}$ has the following expression
\begin{equation}\label{eq:transfer_matrix_dt}
T_{s_i} = \exp[-2 i \Delta t ( \omega L_z+ s_i L_x)],
\end{equation}
where $L_i$ are the generators of $SO(3)$, $(L_i)_{jk} = - i
\epsilon_{ijk}$, satisfying the commutation relations $[L_i,L_j] = i
\sum_k\epsilon_{ijk} L_k$.  The transfer matrix for a $n$-step evolution
may be written as 
\begin{equation}\label{eq:transfer_matrix}
	T = \braket{x_f|\Gamma^n|i_f},
\end{equation} 
where $\ket{x_f} = \frac 1 {\sqrt 2} (\ket{+}+\ket{-})$, $\ket{i_f}$ is
the initial distribution of the states of the fluctuator (in our case
the two states are equiprobable, $\ket{i_f} = \frac 1 {\sqrt 2}
(\ket{+}+\ket{-})$) and $\Gamma$ is the $6 \times 6$ matrix
\begin{equation}
\begin{split}
\Gamma = & \left[(1 - \gamma \Delta t) \id_2 + 
\gamma \Delta t \sigma_1\right] \otimes \id_3 \times \\
&  \exp[-2 i \Delta t ( \omega L_z \id_2 + L_x \sigma_3)],
\end{split}
\end{equation}
where $\times$ denotes a product between $6\times 6$ matrices. The
partial inner product in Eq. \eqref{eq:transfer_matrix} is done on the
two degrees of freedom of the fluctuator and the result is a $3 \times
3$ matrix.  \par
In the continuous limit $\Delta t \rightarrow 0$, Eq.
\eqref{eq:transfer_matrix} becomes
\begin{equation}\label{eq:transfer_matrix2}
	T = \braket{x_f|\exp(-t P)|i_f},
\end{equation}
where
\begin{equation}\label{pdef} P = (\gamma - \gamma \sigma_1)\otimes \id_3
- 2 i \omega \id_2 \otimes L_z - 2 i \sigma_3 \otimes L_x.
\end{equation} 
The problem is now cast into the diagonalization of the
$6\times 6$ matrix $P$. The eigenvalues $\mu_i$, $\eta_i$, $i=1,2,3$, of
$P$ satisfy  the two equations
\begin{align}\label{eq:eig1}
\mu^3 + 2 \gamma \mu^2 + 
4(1+\omega^2) \mu + 8\omega^2 \gamma &= 0 \\ \label{eq:eig2}
\eta^3 + 4 \gamma \eta^2 + 4(1 
+ \gamma^2 + \omega^2)\eta+ 8 \gamma & = 0.
\end{align}
We notice that we can linearly transform one equation into the other by
substituting $\nu = -\mu - 2\gamma$. The inverse of the real parts of
these eigenvalues give the decay rate of the Bloch vector components,
while the inverse of the imaginary parts give the periods of
oscillations.  The matrix elements of $T$ are reported in the Appendix
for reference.
%%%%
\begin{figure}[t]
\includegraphics{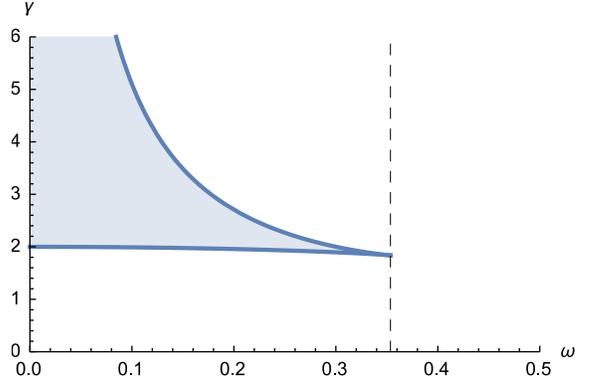}
\caption{Eigenvalues of the operator $P$, see Eq. (\ref{pdef}), as a
function of the qubit energy $\omega$ and the spectral width $\gamma$. 
In the shaded region all the eigenvalues determined by Eqs.
\eqref{eq:eig1} and \eqref{eq:eig2} are real, i.e. there are no
oscillating terms in the transfer matrix. The vertical dashed
line is at the threshold value $\omega = (2\sqrt{2})^{-1}$.}
\label{fig:discriminant}
\end{figure}
%%%%
\par
In the limiting cases of $\gamma$ much greater or smaller than the other
two parameters we are able to obtain analytic expressions for the
eigenvalues.  When $\gamma \gg \omega$, i.e. we are in the fast-noise
regime, we find that the greatest decay time is 
\begin{equation}
	T = \gamma,
\end{equation}
while the oscillation frequency is $\omega$, independently of $\gamma$.
In the opposite limiting case, $\gamma \ll \omega$, we find that the
longest decay time is 
\begin{equation}
T = \begin{cases}
\gamma^{-1}(1+\omega^2) & \text{if }\omega > 1 / \sqrt{2} \\
\frac{1}{2}\gamma^{-1}(1+1/\omega^2) & \text{if }\omega < 1 /\sqrt{2}
\end{cases},
\end{equation}
while the oscillation frequency is instead $\sqrt{1+\omega^2}$.  In the
intermediate region, by studying the discriminant of Eq.
\eqref{eq:eig1}, we find that for $\omega < 1 / (2 \sqrt{2})$ there is a
region of values of $\gamma$ for which the eigenvalues are all real,
i.e. there are no oscillations. This region, shown in Fig.
\ref{fig:discriminant}, is bounded from below and above, respectively,
by the two positive solutions $\gamma_{1,2}$ of 
\begin{equation}
4 \omega ^2 \gamma^4 + \left(8 \omega ^4-20 
\omega ^2-1\right)\gamma^2 + 4 \left(\omega ^2+1\right)^3 = 0.
\end{equation}
For $\omega \rightarrow 0$, $\gamma_1 \rightarrow 2$ and $\gamma_2
\rightarrow \infty$, so we recover the transition between fast and slow
RTN that is visible in the dephasing case \cite{Bordone2012}. In fact,
by letting $\omega \rightarrow 0$ we are implying that the energy gap
between the levels of the qubit is far away from the typical frequencies
of the noise.
A sharp transition between the two regimes is not visible by looking at
the time evolution of the Bloch components because the imaginary
components tend to zero as the parameters get close to the region, and
thus the period of oscillation becomes much larger than the
characteristic decay time.
%%%
\begin{figure}[t!]
\includegraphics{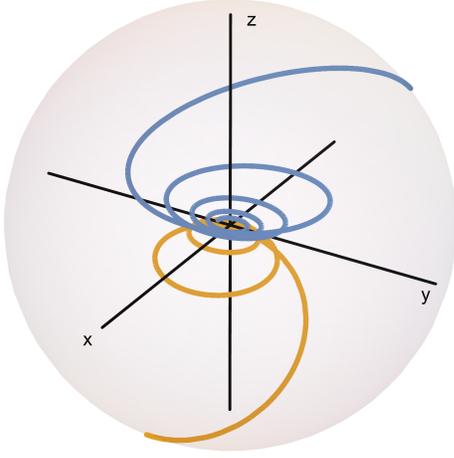}
\caption{Dynamical trajectories in the Bloch sphere for a single 
qubit affected by RTN with $\gamma = 1/2$ and $\omega = 1$ and 
for different initial preparations. The initial state is 
represented by the Bloch vector $\frac 1 {\sqrt 3} (-1, 1, 1)$ 
for the blue trajectory, and by $\frac 1 {\sqrt{2}}(1,0,-1)$ for 
the orange trajectory. The asymptotic state is the maximally 
mixed state, with Bloch vector $(0,0,0)$.}
\label{fig:trajectory_1q}
\end{figure}
\par
In Fig.~\ref{fig:trajectory_1q} we show the dynamical trajectories 
in the Bloch sphere for two different initial preparations. The asymptotic 
state is the maximally mixed state, with Bloch vector $(0,0,0)$.
%%%
\subsection{Transfer matrix for the two-qubit case} % (fold)
\label{sub:transfer_matrix_for_the_two_qubit_case}
The transfer matrix method can be generalized to the two-qubit dynamics 
for both the relevant, and opposed, scenarios
of independent environments and of a common environment.
The generalization of the Bloch vector to the two-qubit 
case is a $15$-component vector defined as follows
\begin{equation}\label{eq:generalized_Bloch_vector}
\mathbf{n}_2 = (\mathbf{a},\mathbf{b},c_{11},c_{12},c_{13}
,c_{21},c_{22},c_{23},c_{31},c_{32},c_{33}),
\end{equation}
where $\mathbf{a} = (a_1,a_2,a_3)$, $\mathbf{b} = (b_1,b_2,b_3)$, 
and $c_{ij}$ are the elements of a $3\times 3$ matrix $C$. The 
two-qubit density matrix may be written as 
\begin{align}\label{eq:Bloch_relation}
\rho = \frac14\, \id_4 &+ \frac 14 \sum_{i=1}^3 \left(a_i \sigma_i 
\otimes \id_2 + b_i\id_2 \otimes \sigma_i \right) \notag \\ 
&+ \frac 14 \sum_{i,j=1}^3 c_{ij} \sigma_i\otimes\sigma_j\,,
\end{align}
where $\mathbf{a}$ and $\mathbf{b}$ are the Bloch vectors of the 
marginals, i.e. of $\rho_1 = \Tr_2(\rho)$ and $\rho_2 = \Tr_1(\rho)$, 
respectively. The action of a unitary transformation on $\rho$ 
corresponds to the action of a real orthogonal transfer matrix 
$T_2$ on $\mathbf{n}_2$. Let us now derive the transfer matrix for
common and independent environments.
%%%
\subsubsection{Common environment}
In the case of a common environment, one can easily see that, when the
common fluctuator is in the state $s_i=\pm 1$, the two-qubit transfer
matrix has the following block-diagonal form: \begin{equation}
T_2 (s_i) = \begin{pmatrix}
T_{s_i} & 0 & 0 \\
0 & T_{s_i} & 0 \\
0 & 0 & T_{s_i} \otimes T_{s_i}
\end{pmatrix},
\end{equation}
where $T_{s_i}$ was defined in Eq.~\eqref{eq:transfer_matrix_dt}. If we
extend the derivation done in the previous subsection for a single
qubit, we obtain the following $30 \times 30$ matrix: \begin{equation}
P_2^\ce = (\gamma \id_2 - \gamma \sigma_1)\otimes \id_{15} - 2 i (\omega
\id_2 \otimes Q_z + \sigma_3 \otimes Q_x), \end{equation} where the
$Q_i$s, with $i=x,y,z$, are $15\times 15$ block-diagonal matrices
\begin{equation}
Q_i = \begin{pmatrix}
L_i & 0 & 0 \\
0 & L_i & 0 \\
0 & 0 & L_i \otimes \id_3 + \id_3 \otimes L_i
\end{pmatrix}.
\end{equation}
The ensemble-averaged transfer matrix for $\mathbf{n}_2$ is then
\begin{equation}
T_2^\ce = \braket{x_f|\exp(-t P_2)|i_f},
\end{equation}
where $\ket{i_f}=\ket{x_f} = \frac{1}{\sqrt 2}(\ket{+}+\ket{-})$ and the
partial inner product is again done on the two degrees of freedom of the
fluctuator.
An analytic expression for $T_2^\ce$ cannot be obtained explicitly 
because we
first need to calculate the exponential of $P_2$, i.e. diagonalize it.
However, the exponentiation can be done easily with arbitrary precision
once we substitute numerical values.
%%%
\subsubsection{Indipendent environments}
In the case of independent environments, the transfer matrix is simply
\begin{equation}\label{eq:T_ie}
T_2^\ie = \begin{pmatrix}
T & 0 & 0 \\
0 & T & 0 \\
0 & 0 & T \otimes T
\end{pmatrix},
\end{equation}
where $T$ is defined in Eq. \eqref{eq:transfer_matrix2}. 
The analytic solution for the one- and two-qubit dynamic under RTN
noise has been compared to the numerical simulations, showing excellent
agreement.
\subsection{Properties of the dynamical map} % (fold)
\label{sub:properties_of_the_dynamical_map}
\begin{figure}[t]
  \centering
\includegraphics[width=.48\columnwidth]{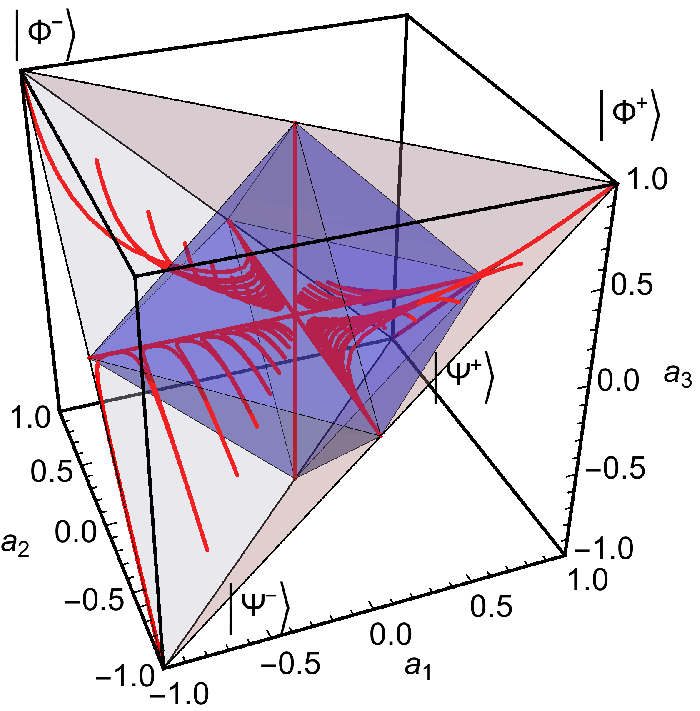} 
\includegraphics[width=.48\columnwidth]{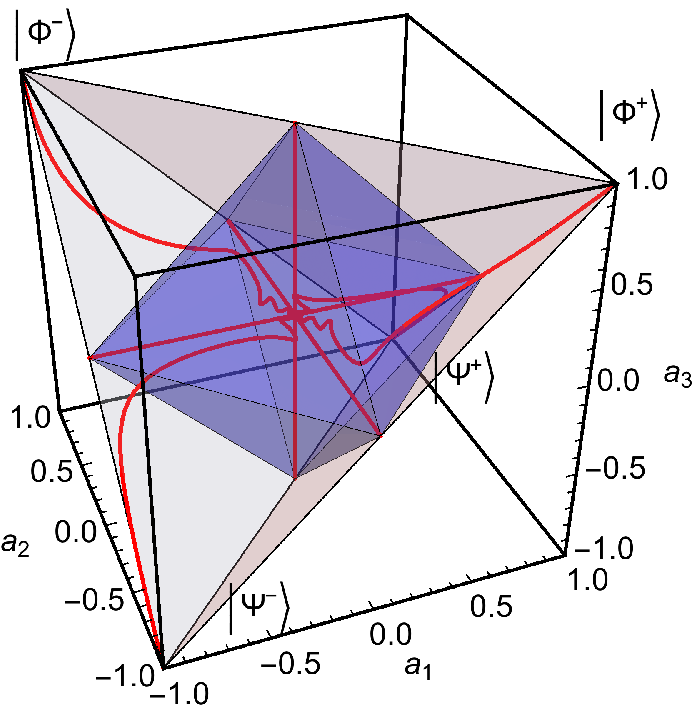} 
\includegraphics[width=.48\columnwidth]{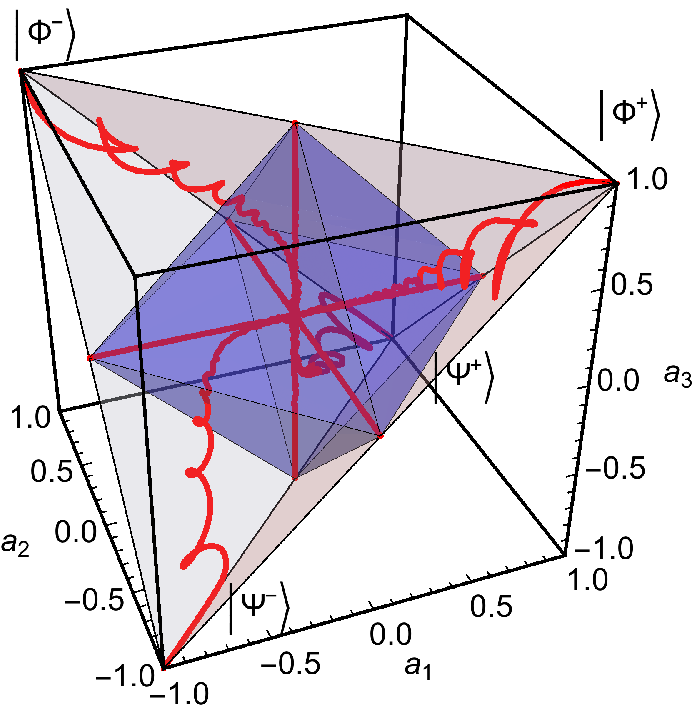}
\includegraphics[width=.48\columnwidth]{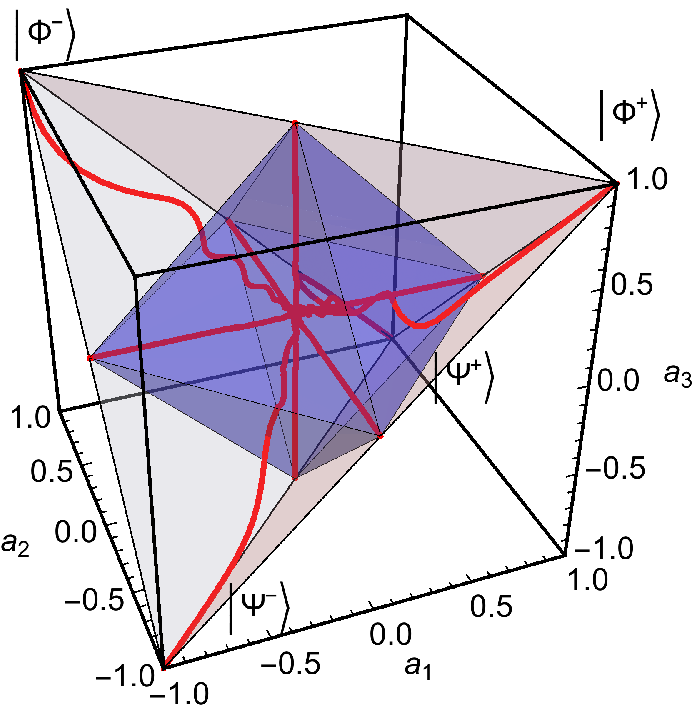} 
\caption{Trajectories of a two-qubit system in the Bell-state
tetrahedron, starting from different initial states, under the influence
of RTN (above) and OU noise (below) with $\gamma = 0.1$ (left) and
$\gamma = 1$ (right), $\omega = 1$. The the dark-blue octahedron is the
set of separable states. We can see that the trajectories converge to
the state in the origin, i.e. the maximally mixed state $\id/4$. The
trajectories, however, get more convoluted for smaller values of
$\gamma$, and, for the RTN noise, one can see that they get in and out
of the set of separable states, and this corresponds to the sudden death
and rebirth of entanglement.} \label{fig:trajectories_2}
\end{figure}
%%%%%
\subsubsection{Maximally mixed marginals} % (fold)
Equation \eqref{eq:T_ie} shows that the two-qubit transfer matrix in the
case of independent environments is block diagonal. The same can be seen
for the matrix $T_2^\ce$. This means that if the initial block vector
has $\mathbf{a} = \mathbf{b} = 0$, i.e. the state has maximally mixed
marginals, then they will be left untouched by the dynamics. This allows
us to apply Luo analytic formula \cite{Luo2008} for quantum discord 
to the evolved state.  
Although we don't have an analytic expression for the dynamics in case
of other kinds of noise, such as Gaussian noise, we can see that the
transfer matrix for an infinitesimal time step is block diagonal as
well. Thus, in general, we can restrict to the set of states with
maximally mixed marginals and use Luo formula  for 
the evaluation of quantum discord.
\par
\begin{figure}[t]
\centering
\includegraphics[width=.48\columnwidth]{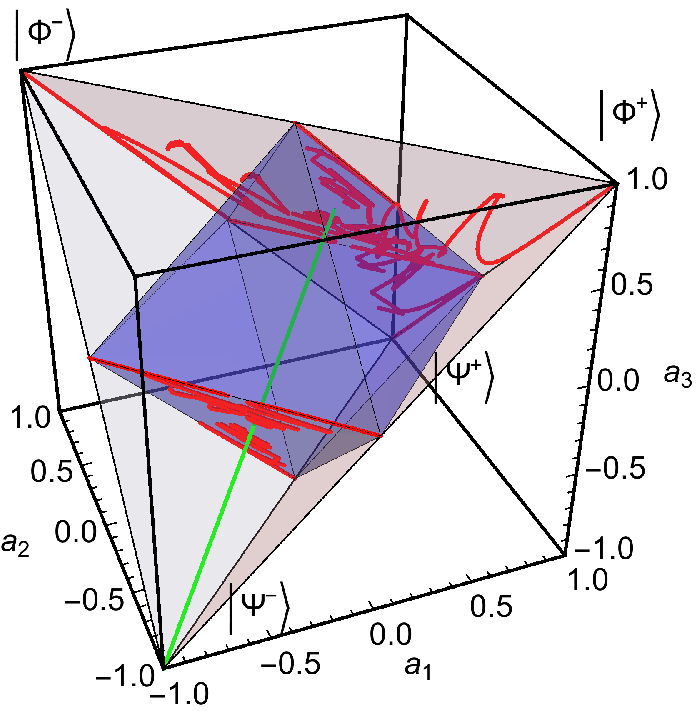}	    
\includegraphics[width=.48\columnwidth]{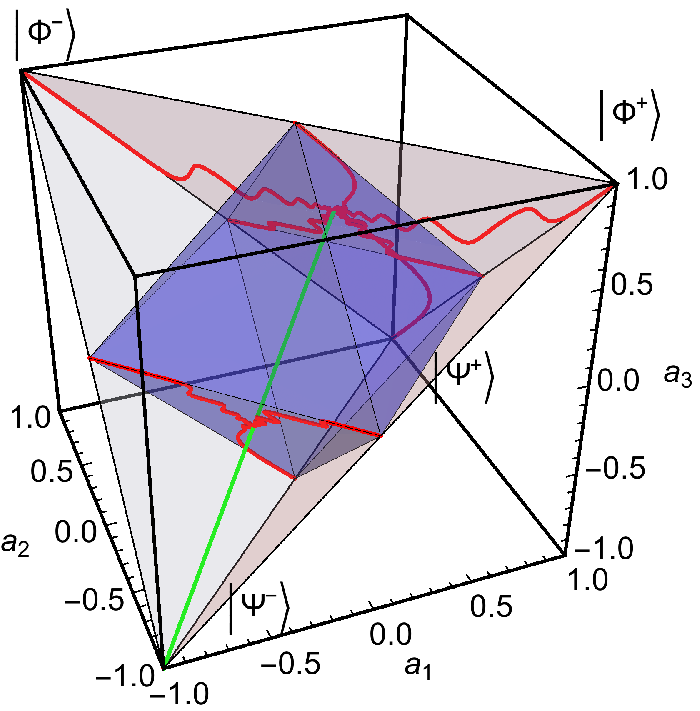}
\caption{Trajectories of the system in the Bell-state tetrahedron when
the qubits interact with a RTN, common environment, $\omega = 1$ for
$\gamma = 1/2$ (left) and $\gamma = 5$ (right). The solid green line 
denotes the set of Werner states, which are the only stable states. The 
trajectories lie on planes that are orthogonal to the green line.  
Similar plots are  obtained for Gaussian noise.}
\label{fig:trajectories_rtn_ce}
\end{figure}

Upon restricting our choice of the initial state to Bell-state mixtures we
are also able to picture the trajectory of the system. In view of the
spectral decomposition theorem, the matrix $C$ of Eq.
\eqref{eq:generalized_Bloch_vector}, if symmetrical, can be diagonalized
by means of an orthogonal matrix, to which correspond two local unitary
operations on the two qubits \cite{Luo2008}. It is straightforward to
check that Bell-state mixtures have a symmetric $C$ matrix. One can also
see analytically that the transfer matrix for the RTN noise with
independent environments, Eq. \eqref{eq:T_ie}, preserves the
symmetric nature of the matrix. The same can be seen numerically for
$T_2^\ce$ and also for Gaussian noise. 
Since all measures of quantum correlations are invariant under local
unitary operations, we can always cast $C$ into its diagonal form, and
represent the two-qubit states with mixed marginals in a tridimensional
space where the coordinates are the eigenvalues of $C$. In this space,
the four Bell states occupy the vertexes of a tetrahedron, as shown in
Figs.~\ref{fig:trajectories_2} and~\ref{fig:trajectories_rtn_ce}. In the
Figures the octahedron of separable Bell-state mixtures is highlighted.
The zero-discord states lie on the axes.
%%%
%%%%
\begin{figure*}[t]
\includegraphics{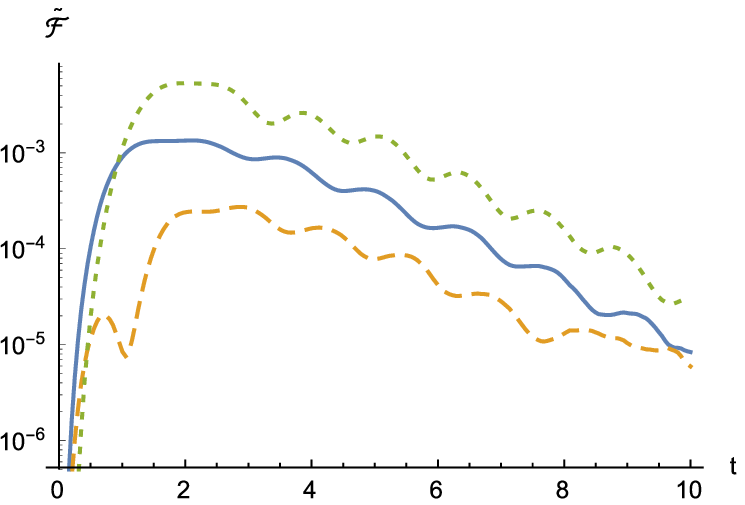}%
\includegraphics{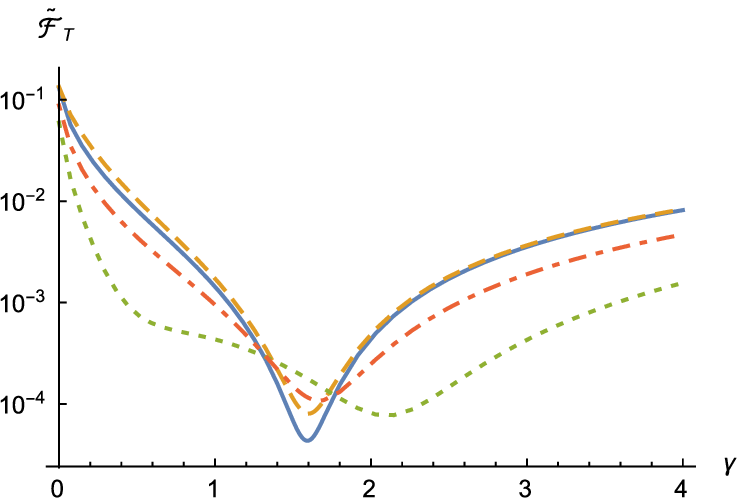}
\caption{Left panel: Logarithm of the fidelity complement 
$\tilde{\mathcal{F}}(t)$ between the state
of a qubit ($\omega = 1$) affected by OU noise (with spectral width 
$\gamma_\ou = 1$) and RTN as a function of (rescaled) time, for 
different values of $\gamma_\rtn$: $\gamma_\rtn=1$ (solid blue), $\gamma_\rtn=1.5$ (dashed orange), $\gamma_\rtn=2$ (dotted green).
The qubit is initially prepared in the state $\mathbf{n} = (1,0,0)$. We
notice that for $\gamma_\rtn = 1$, i.e. when the two noises have the same
spectrum, the dynamics is different. By tuning $\gamma_\rtn$, the
fidelity between the evolved states in the two scenarios may be
increased
by two orders of magnitude. Right panel: the fidelity complement 
$\tilde{\mathcal{F}}_T$ as a function of $\gamma_\rtn$
for $\gamma_\ou = 1$ and for $T = 10$, initial state set to 
$\mathbf{n} = (1,0,0)$ (solid blue), $\mathbf{n} = (0,1,0)$ (dashed yellow), $\mathbf{n} = (0,0,1)$ (dotted green) and $\mathbf{n} = (1,0,1) /\sqrt{2}$ (dot-dashed orange). We can see that the average fidelity depends heavily on the initial state, but that by a suitable choice of $\gamma_\rtn$ we can obtain an average fidelity above $0.9999$.} \label{fig:avg_fidelity}
\end{figure*}
%%%
\subsubsection{Stable states of the dynamics} % (fold)
\label{par:stable_states_of_the_dynamics}
For the single-qubit RTN map the only fixed point is the maximally mixed
state (with the Bloch vector $\vec 0$). This can be seen from the fact
that none of the eigenvalues of $P$ is zero and thus the transfer matrix
doesn't have one as eigenvalue. Figure~\ref{fig:trajectory_1q} shows two
trajectories, both converging to the center of the Bloch sphere.  The
same generalizes immediately to the two-qubit case with independent
environments. The stable state is the maximally mixed state $\rho = \id
/4$.
In the CE case the $P_2$ matrix has the eigenvalue zero. The
corresponding eigenvector is the generalized Bloch vector with
$\mathbf{a}= \mathbf{b} = 0$ and $(c_{ij}) = \id_3$. This means that all
Werner states of the form 
\begin{equation}
\rho^W_p = p \ket{\Phi^-}\bra{\Phi^-} + (1-p) \id/4\qquad p \in [0,1],
\end{equation}
where $\ket{\Phi^-} = 1/\sqrt{2}(\ket{01}-\ket{10})$ is the Bell singlet
state, are stationary states of the dynamics. This can also be seen
because they satisfy the relation $\rho^W_p = (U\otimes U) \rho^W_p
(U^\dagger \otimes U^\dagger)$ for every local unitary $U$ and the CPT
map induced by a common reservoir is a convex combination of unitary
maps of the form $U \otimes U$. Being the zero eigenvalue of $P_2$
non-degenerate, these are the only stable states of the map.
The same results are seen numerically for the Gaussian noise, although
in this case we don't have an analytic expression for the transfer
matrix.	
\par
%%%
\section{Comparison of the dynamics in the presence of 
Gaussian and non-Gaussian noise}
\label{sec:comparison}
In this Section we compare the dynamics induced by Gaussian 
and non-Gaussian RTN noise and discuss their effects on the 
decoherence of quantum correlations of a two-qubit system.
We start by noticing that the spectrum of the noise (or equivalently,
its autocorrelation function) is in general not enough to describe 
the effect of the noise on the qubit, i.e. the dynamics of the qubit 
under the influence of OU noise and RTN with the same spectral width 
and with the same coupling may be, in general, rather different. 
\par%%%
\begin{figure*}[t]
\includegraphics{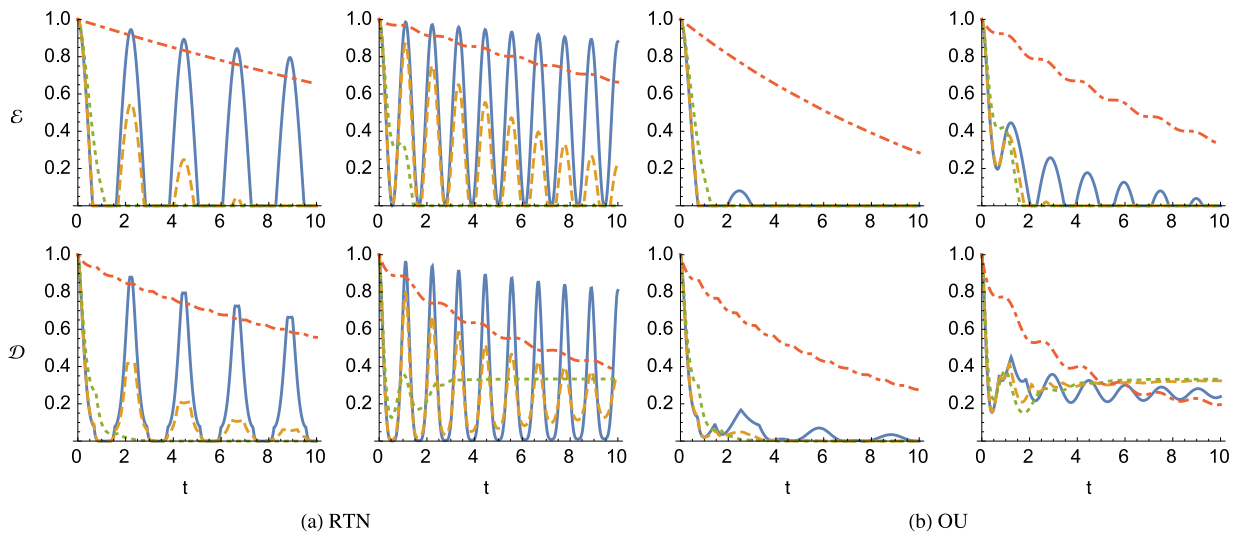}                    
\caption{Negativity $\mathcal{E}$(above) and discord $\mathcal{D}$ (below) as functions of time for
a two-qubit system initially prepared in the Bell state $\ket{\Psi^+} =
1/\sqrt{2}(\ket{00}+\ket{11})$ subject to (a) RTN and (b) OU noise, with
independent (left) and common (right) environment ($\omega = 1$). The
blue solid line is for $\gamma = 10^{-2}$, the yellow dashed one is for
$\gamma = 10^{-1}$, the dotted green is for $\gamma =1 $ and the orange
dot-dashed line for $\gamma = 100$. For both noises, for smaller values
of $\gamma$, quantum correlations oscillate heavily, with sudden deaths
and rebirths of entanglement. The effect is more evident for the RTN.
The frequency of oscillations doubles in the common-environment case.
For higher values of $\gamma$, the correlations decay, possibly with
small oscillations, and entanglement dies suddenly.}
\label{fig:quantum_correlations}
\end{figure*} 
In order to compare quantitatively the dynamics of the system in the
presence of the two kind of noise we introduce the  fidelity complement 
\begin{equation}\label{eq:tilde_fidelity}
\tilde{\mathcal{F}}(t) = 1-\mathcal{F}(\rho_\ou(t),\rho_\rtn(t)),
\end{equation} 
where $\mathcal{F}(\rho_\ou(t),\rho_\rtn(t))$ is the fidelity between
the state of a single qubit affected by RTN and the state of a qubit
affected by OU, assuming that the two kinds of noise have 
the same coupling and spectral width.
When this quantity is zero, the two states are identical. 
In the left panel of Fig.~\ref{fig:avg_fidelity} we show the fidelity 
complement as a function of of time. We can see
that $\tilde{\mathcal{F}}(t)$ is not vanishing when the two noises have
the same autocorrelation time. However, upon changing $\gamma$, we 
can reduce its value of three orders of magnitude. In the right panel, 
we show that the average of $\tilde{\mathcal{F}}(t)$ over the
interaction  time, i.e.
\begin{equation}\label{eq:fidelity_int_tilde}
\tilde{\mathcal{F}}_T = \frac 1 T \int_0^T \tilde{\mathcal{F}}(t) dt,
\end{equation}
can be driven very close to zero by a suitable choice of the parameter
$\gamma$. It should be noticed, however, that the optimal value of the
parameters does depend on the frequency of the qubit, on the parameters
of the OU noise, and also on the initial state of the qubit, as
it is apparent upon looking at the 
right panel of Fig.~\ref{fig:avg_fidelity}. 
We thus conclude that the effects of
non-Gaussian noise on qubits cannot be trivially mapped to that of
Gaussian noise and viceversa. This means that the spectrum alone is not
enough to characterize the effect of the
noise on the qubit systems. On the other hand, the effect of the two
noises is qualitatively similar and the dynamics under the effect of one
kind of noise may be {\em simulated} with high (quantum) 
fidelity with the other kind of noise by suitably tuning the parameters.
\par
In Fig.~\ref{fig:quantum_correlations} we show how the negativity and
quantum discord evolve in time for the two models of noise for various
values of the spectral width $\gamma$. The initial state is a pure Bell
state. For both noises, we can identify two working regimes. 		
In the first one, for small $\gamma$ (slow noise), quantum correlations oscillate
heavily and there are sudden deaths and rebirths of entanglement. This
can be seen in the top left diagram of Fig.~\ref{fig:trajectories_2}:
the trajectory of the system repeatedly goes in and out the octahedron
of separable states. The frequency of oscillations depends on $\omega$
and is doubled if the two qubits are affected by a common environment.
In the second regime (large $\gamma$, i.e. fast noise), the 
correlations decay to zero, with sudden death
of entanglement and with oscillations. The time constant of the decay is
roughly inversely proportional to $\gamma$, i.e. the decay is slower for
very fast noise. In the common-environment case, we notice that the
discord does not vanish in time. The reason is that the stable state of
the dynamics does not lie in the set of states with zero discord (cf.
Fig.~\ref{fig:trajectories_rtn_ce}).
%%%%%%
\section{Comparison of non-Markovianity measures} % (fold)
\label{sec:comparison_of_non_markovianity_measures}
In this Section we evaluate the trace-distance-based BLP measure and 
the entanglement-based RHP measure for the single-qubit map with 
RTN noise. As a comparison, we recall that the dephasing map with 
RTN noise \cite{Benedetti2014c} is Markovian in the regime of fast 
noise, i.e. when $\gamma > 2$, while it is non-Markovian in the other 
regime (slow noise).
\par
\begin{figure*}[t]
\includegraphics{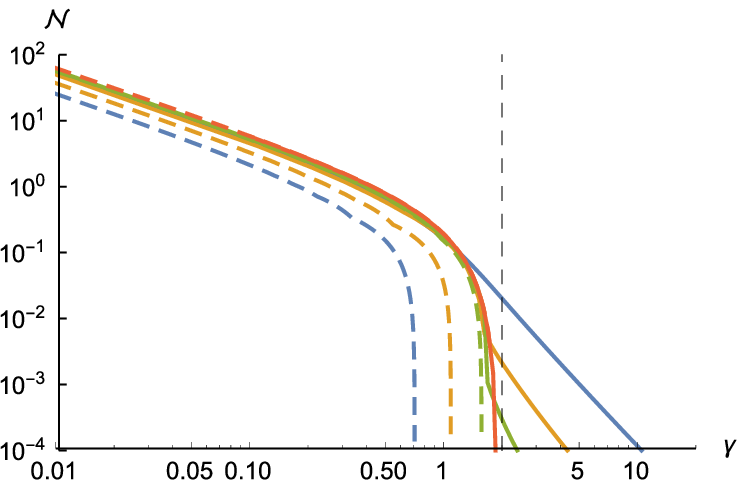}%
\includegraphics{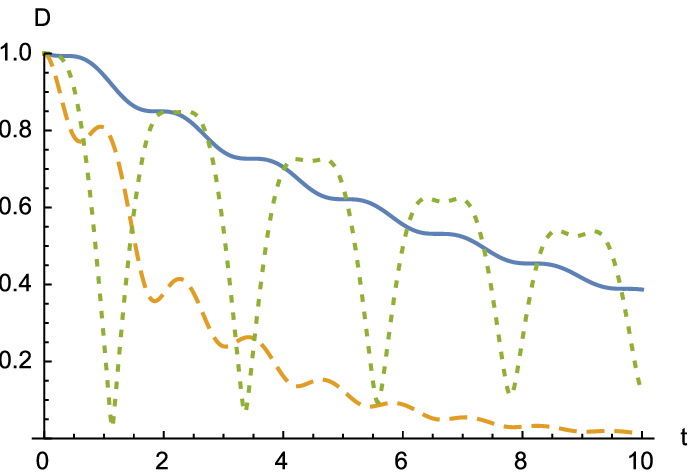}
\caption{Left panel: Log-log plot of the non-Markovianity measures $\blp$ (solid) and $\rhp$ (dashed) as functions of the spectral width $\gamma$ for a qubit subjected to RTN noise, for $\omega=1$ (blue), $\omega=(2\sqrt{2})^{-1}$ (yellow),
$\omega = 0.1$ (green) and $\omega = 0.01$ (orange). The two measures decrease
monotonically for increasing $\gamma$. There is a threshold value for
$\gamma$ (that depends on $\omega$) above which the RHP measure is zero.
The BLP measure, instead, is always non-zero and vanishes for $\gamma
\rightarrow \infty$, i.e. when $K(t-t')\sim \delta(t-t')$. For small
$\gamma$, both measures are proportional to $1/\gamma$. For small $\omega$ (orange
line), we recover the results obtained for the dephasing, with both
measures vanishing at $\gamma = 2$ (vertical dashed line).
Right panel: Trace distance $D$ between the pairs of states that maximize Eq. \eqref{eq:blp} as a function of time for $\omega = 1$ and $\gamma = 0.1$ (dotted green), $\gamma = 1$ (dashed orange), $\gamma = 10$ (solid blue). We see that the trace distance oscillates in time: in the intervals in which it increases the map is not divisible. The oscillations get smaller for higher $\gamma$: they are barely noticeable in the plot for $\gamma = 10$.}
\label{fig:nonmarkovianity}
\end{figure*}

For the BLP measure, our numerical results show that the pair of optimal
states lies on the equator of the Bloch sphere (i.e. $n_z = 0$),
independently on the parameters of the noise. A numerical optimization
over the azimuth angle is still in order for computing the measure.
The optimal angle depends on the two parameters $\gamma$ and $\omega$
and the dependence is sometimes not smooth. We found that the two 
measures are in disagreement, i.e. the BLP measure is always non-zero 
and is vanishing for $\gamma \rightarrow \infty$, whereas the RHP measure 
is vanishing for $\gamma$ greater than a certain threshold. This is 
shown in Fig.~\ref{fig:nonmarkovianity}, where the two measures are 
calculated for a range of values of the switching rate $\gamma$ and for 
different values of $\omega$. From Fig.~\ref{fig:nonmarkovianity} we 
can see that both measures depend approximately on $1/\gamma$ at 
small $\gamma$. While the RHP measure suddenly goes to zero for 
$\gamma$ above a certain threshold value, which depends on $\omega$, 
the BLP measure only vanishes asymptotically. The BLP measure appears 
to be independent of $\omega$ at small values of $\gamma$.
%%%
%%%
We recall that the two measures only pose a sufficient condition for the
dynamical map to be non-divisible, i.e. non-Markovian. The RHP measure
fails to capture the non-Markovian behavior of the map because the
trajectory quickly enters the set of separable states, as one can see
from Fig.~\ref{fig:trajectories_2}.  On the other hand, the BLP measure
is always non-zero, meaning that the map is non-Markovian, unless we let
$\omega \rightarrow 0$. In this case we approach the dephasing limit,
and the BLP and RHP measures coincide and vanish at $\gamma = 2$
\cite{Benedetti2014c}.  This is shown in the left panel of Fig.~\ref{fig:nonmarkovianity}
for $\omega = 0.01$ (green line). For non-vanishing $\omega$, the
non-Markovianity measure vanishes for high values of $\gamma$, as one
can expect, since the stochastic process that models the noise tends to
the Markovian limit, i.e. when $K(t-t')\sim \delta(t-t')$.
\par
In the right panel of Fig.~\ref{fig:nonmarkovianity} we show, for different values of $\gamma$, the behavior of the trace distance $D(t)$ between the pair of states that maximize the integral in Eq. \eqref{eq:blp}. For smaller values of $\gamma$, the oscillations are very pronounced. When $\gamma$ increases, the oscillations become less appreciable ($D(t)$ seems to decay monotonically in the plot for $\gamma = 10$, solid blue line), but derivative of the trace distance is always positive in the first oscillation.
\par
Given the need to optimize over an angle, and the need to reach very
long evolution times in order to capture all the oscillations in the
trace distance, evaluating the BLP measure for the Gaussian noise is
practically unfeasible. However, initial pairs of states can be found
for which the trace distance do not decay monotonically for a very wide
range of values of $\gamma$, and this allows us to conclude that also
Gaussian transverse noise with a Lorentzian spectrum induces
non-Markovian quantum dynamics on qubits.
%%%
\section{Conclusions} % (fold)
\label{sec:conclusions}
In this paper we have addressed the dynamics of open single- and
two-qubit systems evolving in a classical fluctuating environment
described either by Gaussian or non-Gaussian transverse noise, i.e.
characterized by a noise spectrum that includes the characteristic
frequency of the qubit(s). In the two-qubit case we have considered
both the interaction with separate and common environments.
\par
We have analyzed in detail the properties of the quantum map and the
dynamics of quantum correlations, also comparing the effects of the two
kinds of noise and discussing the stable states of the dynamics. 
Our results indicate that while non-Gaussian noise leads
to peculiar features that are not present in the Gaussian noise case,
there are regions of the parameter space in which the two noises produce
very similar effects on the dynamics of the qubit, i.e. the dynamics
is determined by the noise spectrum of the environment rather than its
statistics. This means that while in general the spectrum alone is not enough 
to characterize the effect of the noise, the dynamics under the effect of one
kind of noise may be {\em simulated} with high quantum fidelity with the other
kind of noise by suitably tuning the parameters.
\par
Upon studying in details the dynamics of an initially maximally entangled
state we have identified, for both kind of noise, two different 
working regimes. In the first one, when the spectral width of the noise 
$\gamma$ is small, quantum correlations oscillate
heavily and there are sudden deaths and rebirths of entanglement. 
The frequency of oscillations depends on $\omega$
and is doubled if the two qubits are affected by a common environment.
In the second regime, the correlations decay to zero, with sudden death
of entanglement and with oscillations. The time constant of the decay is
roughly inversely proportional to $\gamma$, i.e. the decay is slower for
very fast noise. In the common-environment case, the discord does not 
vanish in time. The reason is that the stable state of
the dynamics does not lie in the set of states with zero discord.
\par
Finally, we have shown that the quantum map is always non-Markovian
(contrarily to what happens for a dephasing dynamics in the presence 
of the same kind of
noise) and we quantified the non-Markovianity with the BLP measure for
the RTN. We also highlighted the discrepancy between the BLP measure and
the RHP measure based on entanglement, which fails to capture the
non-Markovianity of the dynamical map for a region of the parameters. 
%%%
\begin{acknowledgments}
The authors thank C. Benedetti for useful discussions. This work has 
been supported by EU through the Collaborative Project QuProCS (Grant 
Agreement 641277) and by UniMI through the H2020 Transition Grant 
15-6-3008000-625. 
\end{acknowledgments}
%%%
\appendix
\section{Transfer matrix elements}
Here we write explicitly the nonzero elements of the $3\times 3$
transfer matrix $T$ of Eq.~\eqref{eq:transfer_matrix2}. Here $\mu_i$ and
$\eta_i$ are the solutions of Eqs.~\eqref{eq:eig1} and \eqref{eq:eig2}.
%%%
\begin{widetext}
\begin{align}
T_{11} =&  \frac{e^{\mu _2 t} \left[\mu _1 \mu _3 \left(1-2 \omega ^2\right)-2 
\omega ^2 \left(2 \gamma ^2+\gamma  \mu _2-4 \omega ^2\right)\right]}{4 
\left[1-\omega ^2 \left(2 \gamma ^2+\omega ^2\right)\right]+2 \gamma  
\mu _2 \left(1-6 \omega ^2\right)+\mu _2^2 \left(1-5 \omega ^2\right)}
+\frac{e^{\mu _3 t} \left[4 -4 \omega ^2 \left(\gamma ^2+1\right)+2 
\gamma  \mu _3 \left(1-3 \omega ^2\right)+\mu _3^2 \left(1-2 \omega ^2
\right)\right]}{4 \left[1-\omega ^2 \left(2 \gamma ^2+\omega ^2\right)
\right]+2 \gamma  \mu _3 \left(1-6 \omega ^2\right)+\mu _3^2 \left(1-5 
\omega ^2\right)} 
\notag \\
&+ \frac{e^{\mu _1 t} \left\{2 \gamma  \mu _3 
\omega ^2+\mu _2 \left[2 \gamma  \omega ^2+\mu _3 
\left(1-2 \omega ^2\right)\right]+8 \omega ^4\right\}}
{4 \left[1-\omega ^2 \left(2 \gamma ^2+\omega ^2\right)
\right]+2 \gamma  \mu _1 \left(1-6 \omega ^2\right)+\mu _1^2 
\left(1-5 \omega ^2\right)}
\end{align}
%%%
\begin{align}
T_{12} = & \frac{\omega e^{\mu _2 t} \left[4 \omega ^2 
\left(3 \gamma +\mu _2\right)-\gamma  \mu _1 \mu _3\right]}{4 
\left[1-\omega ^2 \left(2 \gamma ^2+\omega ^2\right)\right]
+2 \gamma  \mu _2 \left(1-6 \omega ^2\right)+\mu _2^2 
\left(1-5 \omega ^2\right)} 
+
\frac{\omega e^{\mu _1 t} \left[4 \omega ^2 \left(\mu _3-\gamma 
\right)+\mu _2 \left(\gamma  \mu _3+4 \omega ^2\right)\right]}{8 
\gamma ^2 \omega ^2-4-2 \gamma  \mu _1 \left(1-6 \omega ^2\right)-
\mu _1^2 \left(1-5 \omega ^2\right)+4 \omega ^4} 
\notag \\ 
&+ \frac{\omega e^{\mu _3 t} \left[\mu _3 \left(2 \gamma ^2+
\gamma  \mu _3-4 \omega ^2\right)+4 \gamma  \left(1-2 \omega ^2
\right)\right]}{8 \gamma ^2 \omega ^2 - 4 -2 \gamma  \mu _3 
\left(1-6 \omega ^2\right)-\mu _3^2 \left(1-5 \omega ^2\right)+4 \omega ^4}
= - T_{21}
\end{align}
\begin{align}
T_{22} = & 2 \gamma  \omega ^2 \left\{\frac{e^{\mu _1 t} \left[\mu _2 
\left(\gamma -\mu _3\right)+\gamma  \mu _3+4 \left(1+\omega ^2\right)
\right]}{\gamma  \left\{4 \left[1-\omega ^2 \left(2 \gamma ^2+
\omega ^2\right)\right]+2 \gamma  \mu _1 \left(1-6 \omega ^2\right)+
\mu _1^2 \left(1-5 \omega ^2\right)\right\}} 
\right.  
\notag\\
&
+ 
\frac{e^{\mu _2 t} \left(2 \gamma ^2+\gamma  \mu _2-4 +\mu _1 \mu _3-4 
\omega ^2\right)}{\gamma  \left[8 \gamma ^2 \omega ^2-4 -2 \gamma  
\mu _2 \left(1-6 \omega ^2\right)-\mu _2^2 \left(1-5 \omega ^2\right)
+4 \omega ^4\right]} 
\notag\\
&
- \left. 
\frac{\left(2 \gamma +\mu _3\right) e^{\mu _3 t}}{4 
\left[1+2 \omega ^2 (1-\gamma^2)+\omega ^4\right]+2 
\gamma  \mu _3 \left(1-2 \omega ^2\right)+\mu _3^2 
\left(1+\omega ^2\right)}\right\}
\end{align}
%%%
\begin{align}
T_{33} = & 2 \omega ^2 \left\{\frac{\left(8 1-\eta _1 \eta _3\right) 
e^{\eta _2 t}}{8 \left[1-\omega ^2 \left(\gamma ^2+\omega ^2\right)
\right]+4 \gamma  \eta _2 \left(1-4 \omega ^2\right)+2 \eta _2^2 
\left(1-5 \omega ^2\right)} \right. \notag \\
&+ \frac{\left(8 -\eta _2 \eta _3\right) e^{\eta _1 t}}{8 \left[1-
\omega ^2 \left(\gamma ^2+\omega ^2\right)\right]+4 \gamma  \eta _1 
\left(1-4 \omega ^2\right)+2 \eta _1^2 \left(1-5 \omega ^2\right)}
\notag \\
& 
+ \left. \frac{e^{\eta _3 t} \left[4 \gamma  \eta _3+4 \left(\gamma ^2
-1+\omega ^2\right)+\eta _3^2\right]}{8 \left(\gamma ^2 \omega ^2-1+
\omega ^4\right)-2 \eta _3 \left[2 \gamma  \left(1-4 \omega ^2\right)+
\eta _3 \left(1-5 \omega ^2\right)\right]}\right\}
\end{align}
\end{widetext}
\bibliography{qtrxlib.bib}
%%%%
\end{document}